\documentclass[prl,twocolumn,superscriptaddress]{revtex4}
\usepackage{graphicx}
\usepackage{amssymb}
\usepackage{amsmath}
\usepackage{xspace}

\begin{document}
\title{Correlation-induced metal insulator transition in a
two-channel fermion-boson model}
\author{G. Wellein}
\affiliation{
Regionales Rechenzentrum Erlangen, Universit\"at Erlangen-N\"urnberg,  
91058 Erlangen, Germany }
\author{H.~Fehske}
\affiliation{
Institut f\"ur Physik, Ernst-Moritz-Arndt-Universit{\"a}t
Greifswald, 17489 Greifswald, Germany }
\author{A.~Alvermann}
\affiliation{
Institut f\"ur Physik, Ernst-Moritz-Arndt-Universit{\"a}t Greifswald, 
17489 Greifswald, Germany }
\author{D. M. Edwards}
\affiliation{
Department of Mathematics, Imperial College London, 
London SW7 2AZ, United Kingdom
}
\begin{abstract}
We investigate charge transport within some background medium
by means of an effective lattice model with a novel form of
fermion-boson coupling. The bosons describe fluctuations
of a correlated background. By analyzing groundstate and 
spectral properties of this transport model, we show how a 
metal-insulator quantum phase transition can occur for the 
half-filled band case. We discuss the evolution of a mass-asymmetric
band structure in the insulating phase and establish connections to   
the  Mott and Peierls transition scenarios. 
\end{abstract}
\maketitle

The way a material evolves from a metallic to an insulating
state is one of the most fundamental problems in solid state physics.
Besides band structure and disorder effects, 
electron-electron and electron-phonon interactions are the 
driving forces behind metal-insulator transitions (MITs) in the majority 
of cases. While the so-called Mott-Hubbard MIT~\cite{Mot90} 
is caused by strong Coulomb correlations, the Peierls MIT~\cite{Pe55}  
is triggered by the coupling to vibrational excitations of 
the crystal. Both scenarios are known to compete in 
subtle ways.  

The MIT problem can be addressed by the investigation 
of generic Hamiltonians for interacting electrons and phonons
like the Holstein~\cite{Ho59,BMH98}, Hubbard~\cite{Hu63,Mot90} and 
(quarter-filled) $t$-$J$ models~\cite{MSU98}, 
or combinations of these~\cite{TC03}. 
These models have been proved to describe MIT phenomena for 
the half-filled band case, in particular for one-dimensional 
(1D) systems known to be  susceptible to the formation of 
insulating spin-density-wave (SDW) or charge-density-wave     
(CDW) broken-symmetry groundstates~\cite{BS93}. 
On the metallic side of the MIT, charge transport then takes 
place within a ``background medium'' that exhibits strong correlations,
which anticipate the SDW or CDW on the insulating side. 
In that case, a particle, as it moves, creates 
local distortions of substantial energy in the background. 
These distortions may be parametrized as bosons. They are able to 
relax, with a rate that depends on the system properties 
but also on the proximity to the MIT.  

In order to model such a situation the authors 
recently proposed a simplified transport 
Hamiltonian~\cite{Ed06,AEF07}     
\begin{equation}\label{hem}
H =  H_b - \lambda \sum_i (b_i^{\dagger} + b_i^{}) 
+ \omega_0 \sum_i b_i^{\dagger} b_i^{}\,, 
\end{equation}
where $H_b=  -t_b \sum_{\langle i, j \rangle}  c_j^{\dagger} c_i^{} (b_i^{\dagger}
+ b^{}_j)$ describes the boson-affected nearest neighbor (NN) hopping     
of fermionic particles $(c_i^\dagger)$~\cite{comment1}. 
In \eqref{hem} the particle  
creates a boson $(b_i^\dagger)$ on the site it leaves  
and destroys a boson on the site it enters. Thereby it 
generates a ``string'' of local bosonic fluctuations 
with energy $\omega_0$~\cite{BR70}. Cutting the string, the $\lambda$ term 
allows a boson to decay spontaneously. A unitary transformation 
$b_i \mapsto b_i + \lambda / \omega_0$ 
eliminates the boson relaxation term 
in favor of a free-particle hopping channel,  
$H_f= - t_f \sum_{\langle i, j \rangle}  c_j^{\dagger} c^{}_i$
with $t_f=2\lambda t_b/\omega_0$,  
in addition to the original one. As a result 
\begin{equation}\label{tctm}
H \mapsto H=H_b+H_f+ 
\omega_0 \sum_i b_i^{\dagger} b_i\,,
\end{equation}
and the physics of our model is governed by two parameter
ratios:  the relative strengths of the two transport channels 
$(t_f/t_b)$ and the rate of bosonic fluctuations $(\omega_0/t_b)^{-1}$. 
The model has been solved numerically in the one-particle 
sector and revealed---despite its seeming simplicity---a surprisingly rich
``phase diagram'' with regimes of quasi-free, correlation and fluctuation 
dominated transport~\cite{AEF07}. In this case the spinless 
Hamiltonian~\eqref{tctm} covers basic features of the more complicated 
$t$-$J$, Hubbard or Holstein models in the low doping/density regimes, 
but is much easier to evaluate. 

Whether our two-channel transport model likewise describes a 
quantum phase transition from a metallic to an insulating phase 
at certain commensurate band fillings remained an important but 
open question. The free hopping channel $H_f$ will clearly act 
against any correlation induced charge ordering that might open a 
gap at the Fermi energy $E_F$. Strong bosonic fluctuations, 
i.e. small $\omega_0$, will also tend to destroy CDW order.  
On the other hand, a tendency towards CDW formation at half-filling
is expected for large $\omega_0/t_b$ by perturbative arguments, 
yielding an effective Hamiltonian with nearest-neighbor 
fermion repulsion. In some respects this is evocative of the 
quantum phase transition in the spinless fermion Holstein model, 
which for large phonon frequencies and strong couplings can be mapped 
on the XXZ model~\cite{BMH98}. The XXZ model undergoes a 
Kosterlitz-Thouless transition at the spin isotropy point. 

Further evidence of a quantum phase transition comes from the 
investigation of simplified versions of~\eqref{hem}: (i) a 
coarse-grained model with only one bosonic oscillator for 
the 1D infinite system, and (ii) a 1D model based on a 2-site cluster. 
The exact solution of (i) and an approximate solution of (ii) 
both exhibit quantum phase transitions at $\lambda=0$ between 
two Fermi liquids, with different Fermi surfaces, non-Fermi-liquid 
behavior persisting down to zero temperature at the critical point 
in case (ii)~\cite{Ed06}. 

The MIT is a subtle quantum mechanical problem, however, 
that requires non-approximative investigation schemes. 
Therefore, in this work, we apply unbiased numerical techniques
to study the competition between itinerancy, correlations 
and fluctuations for the 1D half-filled band case,  
using the full model \eqref{hem} without any restrictions.
To this end we employ exact diagonalization in combination 
with kernel polynomial expansion methods, adapted for 
coupled fermion-boson systems~\cite{BWF98}.   
The computational requirements are determined by the 
Hilbert space dimension $D_H = {N\choose N_e}{N+N_b \choose N_b}$, 
where $N$ is the number of lattice sites, $N_e$ counts the fermions, 
and $N_b$ is the maximum number of bosons retained.
Typically we deal with $D_H$ of about $10^{11}$. 

Let us start with the discussion of the photoemission (PE) spectra.
The spectral density of single-particle excitations associated 
with the injection of an electron with wave-vector 
$k$, $A^+(k,\omega)$ (inverse PE), 
and the corresponding quantity for the emission of an electron, 
$A^-(k,\omega)$ (PE), are given by
$A^\pm(k,\omega) 
=  \sum_n 
|\langle \psi_n^\pm|c^\pm_k 
|\psi_0\rangle|^2 \,\delta [\omega\mp\omega_n^\pm] $.
Here $c^+_k=c^\dagger_k$, $c^-_k=c^{}_k$, and $|\psi_0\rangle$ 
is the groundstate in the $N_e$-particle sector 
while $| \psi_n^{\pm}\rangle$ denote the $n$-th 
excited states in the $N_e\pm 1$-particle sectors 
with excitation energies $\omega_n^\pm=E_n^\pm-E_0$.

\begin{figure}[t]
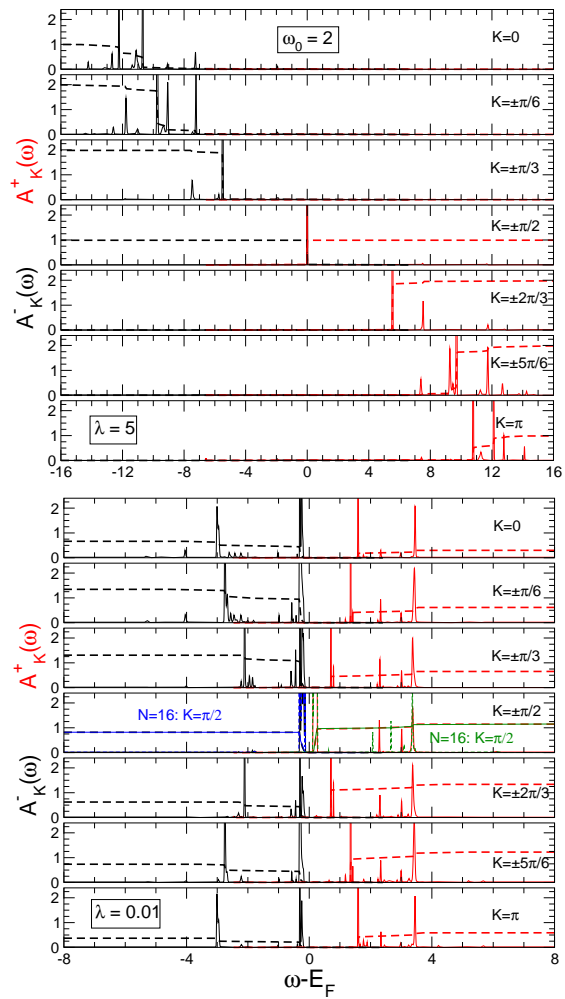

\includegraphics[width=.85\linewidth]{f1a.eps}\\[0.2cm]
\includegraphics[width=.84\linewidth]{f1b.eps}\\
\caption{\label{fig_ak20} 
Photoemission (black) and inverse photoemission spectra (red) 
for the half-filled band case with $\omega_0=2$ at 
$t_f=5$  (upper panels) and $t_f=0.01$ (lower panels),
where $N=12$, $N_b=15$. 
Dashed lines give the integrated spectral weights, e.g.
$S^+_{K}(\omega-E_F)=\int_0^\omega d\omega^\prime A^+_K(\omega^\prime-E_F)$,
where $S_K=S^-_{K}(-\infty)+S^+_{K}(\infty)=1$, and $\sum_KS_K=N$. 
Here and in what follows periodic boundary conditions were used, 
leading to discrete $K(N)$ wave numbers. All energies are measured 
in units of $t_b=1$, and $\omega$ is rescaled with respect to $E_F$.}
\end{figure}
\begin{figure}[h]
\includegraphics[width=.9\linewidth]{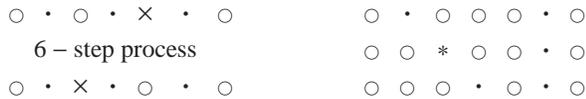}\\[0.5cm]
\caption{\label{fig_2band}Doping a
perfect CDW, states with one particle
removed (left panel) are connected by a  
6-step hopping process (see text), whereas a 
2-step process (right panel) relates states
with an additional particle. 
}
\end{figure}

Figure~\ref{fig_ak20} displays the wave-vector resolved
single-particle spectra in the regime where distortions 
of the background are energy-intensive, i.e. the boson
frequency $\omega_0$ is high. If the free transport channel is dominant 
($t_f=5$ -- upper graph), the occupied (unoccupied) band states, 
probed by PE (inverse PE), give rise to 
an almost particle-hole symmetric absorption spectrum $A^+_K(\omega-E_F)\simeq 
A^-_{K-\pi}(E_F-\omega)$. Thereby the main spectral weight resides 
in the uppermost (lowest) peaks of $A^-_{K}$ ($A^+_K$) in each $K$-sector. 
The corresponding ``coherent'' band structure roughly follows the 
$ -2t_f \cos K $ tight-binding band. Satellites with less spectral weight
occur near the Brillouin zone boundary predominantly, as a result of mixed 
electron-boson excitations with total wave vector $K$.
At $T=0$ the Fermi energy is obtained from 
$\sum_K\int_{-\infty}^{E_F}A_K(\omega)d\omega=N_e=N/2$ 
(half-filling, no spin). We see that there is no gap between 
$A^+_{\pm\pi/2}$ and $A^-_{\pm\pi/2}$ at $E_F$. Moreover the spectral weight
of both peaks is almost one, i.e. a particle injected (removed) 
with $K=K_F=\pm\pi/2$ propagates unaffected by bosonic
fluctuations. The system behaves as an unusual metal.  

If we decrease $\lambda$ ($t_f/t_b$ ratio) at fixed $\omega_0$,
we enter the regime where boson-assisted transport becomes
important (see lower graph of Fig.~\ref{fig_ak20}).
At about  $\lambda_c(\omega_0=2)\simeq 0.1$ a gap opens at 
$K=\pm \pi/2$ in the PE spectra.   
The gap increases as $\lambda<\lambda_c$ gets smaller, but its
magnitude shows no finite-size dependence   
(to demonstrate this we included the $N=16$, $N_b=9$ data  
for $K=\pm \pi/2$). Most notably $E_F$ lies inside the gap region, 
signalising the transition to the insulating state. 
The MIT is correlation induced. Since $\lambda$ is small,  
distortions of the background cannot relax easily. 
Accordingly the band structure is strongly renormalized.
We observe that now  $A^{\pm}_K(\omega-E_F)\simeq A^{\pm}_{\pi-K}(\omega-E_F)$
and expect a perfect doubling of the Brillouin zone for $N\to \infty$. 
Mapping our model~\eqref{hem} for $\omega_0\gg t_f, t_b$ 
to a (XXZ-like) spin model, $S_z\to - S_z$ symmetry is broken, 
reflecting the observed broken particle-hole symmetry: 
The highest occupied states belong to an extremely flat quasiparticle band,
whereas the lowest unoccupied states are much more dispersive~\cite{comment2}. 

\begin{figure}[b]
\includegraphics[width=.85\linewidth]{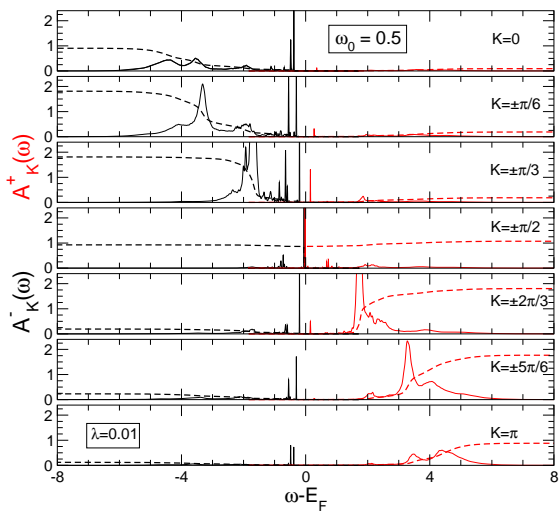}
\caption{\label{fig_ak05} (Inverse) Photoemission   
for $\omega_0=0.5$ and $\lambda=0.01$, i.e. $t_f(\lambda,\omega_0)=0.04$.
Again $N=12$, but now $N_b=15$.
}
\end{figure}

The correlated band structure can be understood
by ``doping'' a perfect CDW state (Fig.~\ref{fig_2band}).
To restore the CDW order a doped hole can be transferred by 
a coherent 6-step process of order ${\cal O}(t_b^6/\omega_0^5)$, 
$|\circ\,\cdot\,\cdot\,\rangle\to
|\ast\,\circ\,\cdot\,\rangle\to
|\ast\,\ast\,\circ\,\rangle\to |\ast\,{\ast \choose \circ}\,\ast\,\rangle
\to |\circ\,\ast\,\ast\rangle\to |\cdot\,\cdot\,\circ\rangle $,
where in steps 1-3, three bosons ($\ast$) are excited, which are consumed
in steps 4-6 afterwards~\cite{AEF07,comment3}. In this process the 
fermion ($\circ$) becomes correlated with the background fluctuations. 
We note that such a coherent hopping process, in which the particle 
propagates and restores the background, exists even for the case 
$\lambda=0$ where transport is fully boson-assisted.
In contrast an additional electron can move by a two-step process 
of order ${\cal O}(t_b^2/\omega_0)$. Consequently the electron band 
is much less renormalized than the hole band, and the mass enhancement 
is by a factor ${\cal O}((t_b/\omega_0)^4)$ smaller.
Note that the mass-asymmetric band structure that evolves here is 
correlation induced.  
 
That the observed MIT is indeed correlation induced is corroborated 
by the weakening and finally closing of the excitation gap if the 
boson energy $\omega_0$ is reduced at fixed $\lambda$ 
(see Fig.~\ref{fig_ak05}). In this way the ability of 
the background to relax is enhanced, fluctuations overcome correlations 
and the system turns back to a metallic state. At the same time
the spectral weight is transferred from the coherent to the incoherent 
part of the (inverse) PE spectra, especially for $K$ away from 
$K_F=\pi/2$ where the lineshape is affected by rather broad bosonic 
signatures. 

The CDW structure of the insulating state becomes apparent
by investigating the particle-particle 
$\chi_{ee}(j)=\frac{1}{N_e^2}\sum_i \langle n_i n_{i+j}\rangle$ and particle-boson 
$\chi_{eb}(j)=\frac{1}{N_e}\sum_i \langle n_i^{} b_{i+j}^\dagger
b_{i+j}^{} \rangle$ correlation functions,  
where $n_i=c_i^\dagger c_i^{}$. 

\begin{figure}
\includegraphics[width=.9\linewidth]{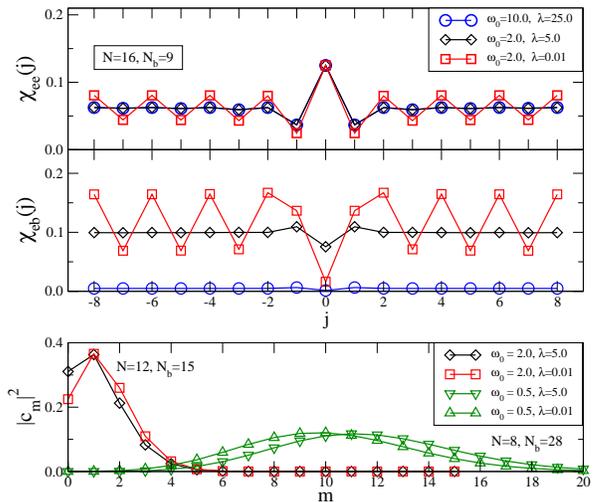}
\caption{\label{fig_gscorr} Particle-particle ($\chi_{ee}$)
respectively particle-boson ($\chi_{eb}$) correlation functions
[upper graph], and weight of the $m$-boson state ($|c_m|^2$)
[lower panel] in the groundstate of the half-filled
1D two-channel fermion-boson model~\eqref{tctm}.}
\end{figure}

In Figure~\ref{fig_gscorr} the even-odd modulation of the charge density 
away from a singled out site $i$ of the first particle
is clearly visible. We note that the  charge structure factor, 
$S_c(\pi)=\tfrac{1}{N^2}\sum_{i,j}(-1)^j\langle 0|(n_i-1/2)(n_j-1/2)|0\rangle$, 
increases by a factor of about two in going from
$\lambda=0.1$ to $\lambda=0.01$ for $\omega_0=2$ (cf. Tab.~\ref{tab})~\cite{comment4}.

In the CDW, where e.g. the even sites are occupied,
every hop of a fermion excites a boson at an even site.
This gives a large contribution to $\chi_{eb}(j)$ 
at even sites 
in addition to NN sites $|j|=1$  
(middle panel). Since the CDW involves only few bosons (see lower panel), 
this is the dominant contribution in first order of $t_b/\omega_0$, and
explains why the boson density is large at sites with large fermion
density, although the hopping term $t_b$ creates bosons at the neighboring 
sites of a fermion.  

The charge oscillations become rapidly suppressed by increasing $\lambda$, 
but there is still a reduced charge density at the particle's 
neighboring sites, which enhances the mobility of the carrier. 
Accordingly the boson density is enlarged  (suppressed) 
at the NN sites (site) of the particle. Clearly
$\chi_{eb}(j)$ is small $\forall |j|$ if $\omega_0\gg t_b,\,t_f$  
because of the high energy cost. As expected the fluctuation 
dominated regime is characterized 
by a large number of boson quanta in the groundstate. The 
position of the maximum in the boson weight function $|c_m|^2$ is shifted
to slightly larger values as $\lambda$ increases, i.e. the 
correlations weaken.

Further information on the groundstate properties can be obtained
from the kinetic energy parts
$E_{kin,f/b} = \langle 0|H_{f/b}|0\rangle$.
Table~\ref{tab} shows that boson-assisted hopping becomes
the major transport mechanism at small $\lambda$. On the 
metallic side of the MIT the creation and annihilation of bosons
opens a coherent transport channel in the regime where
strong correlations persist in the background. 
The Drude weight $D$, obtained from the f-sum rule
$
- D =  
  \tfrac{1}{2}(E_{kin,f}+E_{kin,b}) + 
\int_0^\infty \! \sigma_{reg}(\omega) d\omega$, 
serves as a measure for this coherent transport. 
Here $\sigma_{reg}(\omega) = \sum_{n>0} 
\frac{|\langle n | j |0\rangle |^2}{\omega_n}
   \delta(\omega - \omega_n)$ 
is the regular part of the optical conductivity 
(with current $j = j_f + j_b$).
   
At the MIT point $D$ vanishes for the infinite system. At the same
time the optical gap opens. In the insulating phase the optical
response is dominated by multi-boson emission and absorption processes.
Thus the spectral weight contained in the regular part 
of $\sigma(\omega)$ 
is enhanced.

\begin{table}[t]
\begin{ruledtabular}
\begin{tabular}{ccccccc}
$\lambda$&$\omega_0$&$t_f$&$E_{kin,f}$&$E_{kin,b}$&$D$
&$S_{c}(\pi)$\\[0.1cm]
\hline
5.0&0.5& 20.0 &-149.190  &  -13.915    & 80.336 & 0.0417\\
0.1&0.5& 0.4 & -2.745	  &   -14.878   & 4.054  & 0.0419\\
\hline
5.0&2.0& 5.0 &-37.040  &   -5.291    & 20.290 & 0.0417\\
1.0&2.0& 1.0 &-7.008   &   -6.214    & 5.146  & 0.0425\\
0.1&2.0& 0.1 &-0.568   &   -7.036    &  --    & 0.0561\\
0.01&2.0& 0.01 &-0.023  &   -7.125    &  --    & 0.1056
\end{tabular} 
\end{ruledtabular}
\caption{\label{tab}Groundstate properties
of the two-channel transport Hamiltonian~\eqref{tctm} 
at half-filling ($N=12$, $N_b=15$).}
\end{table}

A small boson frequency allows for large fluctuations in the background,
i.e. many bosons in our model. 
This supports transport via the 6-step process on the one hand but, 
as in the one-particle sector~\cite{AEF07}, 
also limits the mobility of a particle by many 
scattering events. Nevertheless $D$ is expected to stay finite 
even for $\omega_0/t_b\to 0$.  

To summarize, the two-channel transport Hamiltonian, introduced for 
studying the dynamics of charge carriers in a 
correlated/fluctuating medium, has previously only 
been properly analyzed for a single carrier~\cite{AEF07}. 
In this limit the model may capture some of the physics of
2D high-$T_c$ superconducting cuprates~\cite{NT07} or 3D 
colossal magnetoresistive manganites~\cite{WL02}. 
Here we focused on the metal insulator transition problem
at finite particle density, in particular in 1D at half-filling,
which might be of importance, e.g., for the 1D CDW MX chain 
compounds~\cite{BS93}. Since in this case the problem is of the same 
complexity as for the quarter-filled $t$-$J_z$-$J_{\perp}$ or 
spinless fermion Holstein models we make use of elaborate numerical techniques 
in order to avoid uncontrolled approximations. From our finite-cluster study 
we have strong evidence that the model exhibits a quantum phase transition 
from a metallic to an insulating state. The MIT is driven by correlations, 
like the Mott-Hubbard transition, but in our case true 
long-range order develops because a CDW state is formed. 
This might point towards a Peierls transition scenario. 
The Peierls instability, however,  
is most pronounced in the adiabatic limit of small phonon frequencies, 
with many phonons involved in establishing the CDW (lattice dimerization). 
By contrast we find that the CDW groundstate is a few-boson state.
Obviously the system is more susceptible to CDW-formation at 
large boson frequency $\omega_0$ (small transfer amplitude $t_f$),
keeping the boson relaxation $\lambda=t_f\omega_0/2t_b$ fixed.
This is the limit of an effective fermionic system with
(instantaneous) Coulomb repulsion. Recall that as a 
consequence of the correlation-induced CDW state, a band structure with a
very narrow valence and broad conduction band evolves, different
in nature from simple two-band models. 

{\it Acknowledgements.} This work was supported by the Bavarian 
KONWIHR project HQS@HPC and by DFG through SFB 652, B5.


\begin{thebibliography}{21}
\bibitem{comment2}
We found dispersive (flat) absorption bands in the PE (inverse PE) 
spectra as well, but for higher excitation energies and 
with less spectral weight.
\end{thebibliography}
\end{document}